# Software (Re-)Engineering with PSF II: from architecture to implementation

*Bob Diertens*

Programming Research Group, Faculty of Science, University of Amsterdam

*ABSTRACT*

This paper presents ongoing research on the application of PSF in the field of software engineering and reengineering. We build a new implementation for the simulator of the PSF Toolkit starting from the specification in PSF of the architecture of a simple simulator and extend it with features to obtain the architecture of a full simulator. We apply refining and constraining techniques on the specification of the architecture to obtain a specification low enough to build an implementation from.

## 1. Introduction

In this article, as part of ongoing research of the application of PSF (Process Specification Formalism) in the field of software engineering and reengineering, we describe the development of a new implementation of the simulator in the PSF Toolkit. PSF is based on ACP (Algebra of Communicating Processes) [4] and ASF (Algebraic Specification Formalism) [5]. A description of PSF can be found in [21], [22], [11], and [12]. The PSF Toolkit contains among other components a compiler and a simulator that can be coupled to an animation [13]. Animations can either be made by hand or be automatically generated from a PSF specification [14]. Our work is motivated by a range of previous examples of the use of process algebra [3] in the area of architectural description languages (ADL's). We mention *Wright* [2] (based on CSP [18]), *Darwin* [20] (based on the $\pi$-calculus [24]), and *PADL* [7], which is inspired by *Wright* and *Darwin* and focuses on architectural styles.

As case study in previously work we reengineered the compiler from the PSF Toolkit [15]. We developed a PSF specification for the compiler from which we derived a specification of the compiler as a ToolBus application. The ToolBus [6] is a coordination architecture for software applications developed at the CWI (Amsterdam) and the University of Amsterdam. It utilizes a scripting language based on process algebra to describe the communication between software tools. A ToolBus script describes a number of processes that can communicate with each other and with various tools existing outside the ToolBus. The role of the ToolBus when executing the script is to coordinate the various tools in order to perform some complex task. A PSF library of ToolBus internals was developed which was used for the specification of the compiler as ToolBus application. We used this specification to implement the compiler as a real ToolBus application. From the specification we extracted a specification of the architecture of the (reengineered) compiler. By using this architectural specification we built a parallel version of the compiler while reusing specifications and implementations for components of the compiler as it already was configured as a ToolBus application. It was concluded that PSF is useful as aid in software engineering and reengineering, but that also experience should be acquired with starting at the software architecture level and working towards an implementation.

For the development of a new implementation of the simulator in the PSF Toolkit we evaluated the old implementation, which led to the requirements for our new implementation. From these requirements we design an architecture of a simple simulator which we specify in PSF. We extend this with some features for a more complete simulator. For the specification of the architecture we make use of a PSF library especially developed for describing software architectures. This library is really an abstraction of the PSF ToolBus library. We take this architecture specification as a base for the specification of the system design with the use of the PSF ToolBus library, and implement this system. Furthermore we add an history mechanism to the simulator and show the adaptations to be made at the different levels of design. Finally we add animation to the new simulator as was previously done with the old simulator.

## 2. Software Architecture with PSF

A software design consist of several levels, each lower one refining the design on the higher level. The highest level is often referred to as the architecture, the organization of the system as a collection of interacting components. In conventional software engineering processes, the architecture is usually described rather informal by means of a boxes-and-lines diagram. Following a lot of research going on in this area architectural descriptions are becoming more formal, especially due to the introduction of architectural description languages (ADL's). A specification in an ADL can be refined (in several steps) to a design from which an implementation of the system can be built.

In this section we present a PSF library for specifying software architectures or to formalize the boxes-and-lines diagram. With the use of the PSF Toolkit it is possible to generate an animation from the specification which can be brought to live with the simulator of the Toolkit. We also give an example of how to use it.

### 2.1 Specification of the PSF Architecture library

First we define the types for the id's of the components, connections between components, and the data.

```
data module ArchitectureTypes
begin
    exports
    begin
        sorts
            ID,
            CONNECTION,
            DATA
        functions
            _>>_ : ID # ID → CONNECTION
    end
end ArchitectureTypes
```

We could do without the function >> and use just the two id's but now the connection clearly stands out from other terms and therefor makes the specifications easier to read.

We define the primitives for the communication between the components and an quitting action that communicates with the architecture environment.

```
process module ArchitecturePrimitives
begin
    exports
    begin
        atoms
            snd : CONNECTION # DATA
            rec : CONNECTION # DATA
            comm : CONNECTION # DATA

            snd-quit
    end
    imports
        ArchitectureTypes
    communications
        snd(c, s) | rec(c, s) = comm(c, s) for c in CONNECTION, s in DATA
end ArchitecturePrimitives
```

Note that we do not specify a particular kind of connection. In our belief the choice of the kind of connection should not be made on the architecture level, but on a lower level.

We now specify the architecture environment parameterized with the architecture specification.

```
process module Architecture
begin
    parameters
        System
        begin
            processes
                System
        end System
    exports
    begin
        processes
            Architecture
    end
    imports
        ArchitecturePrimitives
    atoms
        rec-quit
        quit
        snd-shutdown
        rec-shutdown
        shutdown
    processes
        ArchitectureControl
        ArchitectureShutdown
    sets
        of atoms
            H = {
                snd(c, s), rec(c, s) | c in CONNECTION, s in DATA
            }
            ArchitectureH = {
                snd-quit, rec-quit,
                snd-shutdown, rec-shutdown
            }
    communications
        snd-quit | rec-quit = quit
        snd-shutdown | rec-shutdown = shutdown
    definitions
        Architecture =
            encaps(ArchitectureH,
                disrupt(
                    encaps(H, System),
                    ArchitectureShutdown
                )
                ‖  ArchitectureControl
            )
        ArchitectureControl =
            rec-quit .
            snd-shutdown
        ArchitectureShutdown = rec-shutdown
end Architecture
```

PSF does not have an action to end all processes. Such an action is really a communication with the environment in which the processes run and this environment has to end all processes. We have specified this behavior with the processes ArchitectureControl as the environment, ArchitectureShutdown to disrupt the running of the processes, and splitting up the actions quit and shutdown in a send and receive part.

*2.2 Example*

As an example of the use of the PSF Architecture library, we specify the architecture of an application in which one component can either send a 'message' to another component and wait for an acknowledgement from that component, or it can send a 'quit' after which the application will be shutdown by the architecture environment.

We first specify a module for the data and id's we use.

```
data module Data
begin
```

```
        exports
        begin
            functions
                message : → DATA
                ack     : → DATA
                quit    : → DATA

                c1 : → ID
                c2 : → ID
        end
        imports
            ArchitectureTypes
    end Data
```

We then specify the system of our application.

```
    process module ApplicationSystem
    begin
        exports
        begin
            processes
                ApplicationSystem
        end
        imports
            Data,
            ArchitecturePrimitives
        atoms
            send-message
            stop
        processes
            Component1
            Component2
        definitions
            Component1 =
                send-message .
                snd(c1 >> c2, message) .
                rec(c2 >> c1, ack) .
                Component1
             +  stop .
                snd-quit
            Component2 =
                rec(c1 >> c2, message) .
                snd(c2 >> c1, ack) .
                Component2
            ApplicationSystem = Component1 ‖ Component2
    end ApplicationSystem
```

And we put it in the architecture environment by means of binding the main process to the System parameter of the environment.

```
    process module Application
    begin
        imports
            Architecture {
                System bound by [
                    System → ApplicationSystem
                ] to ApplicationSystem
                renamed by [
                    Architecture → Application
                ]
            }
    end Application
```

The generated animation of the architecture is shown in figure 1. Here, Component1 has just sent a message to Component2, which is ready to send an acknowledgement back.

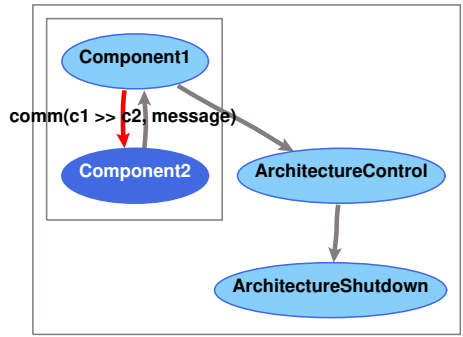

**Figure 1.** Animation of an example architecture

Each box represents an encapsulation of the processes inside the box, and a darker ellipse is a process which is enabled to perform an action in the given state.

The module mechanism of PSF can be used for more complex components to hide the internal actions and sub-processes of a component. With the use of parameterization it is even possible to make several instances of a component.

## 3. Requirements for the Simulator

Although our old simulator from the PSF Toolkit is most satisfactory, we think its implementation can improve a lot. Its interface is outdated and the internal complexity can be lifted from the kernel of the simulator and pushed to separate components and their interaction. We give in this section the requirements for the new simulator without going into much detail. They merely serve to give an idea of what the simulator should be capable of and what we expect from the new design.

### 3.1 Functional Requirements

The functional requirements we list here stem from the functionality of the old simulator. Some features have been left out because they are very seldom used and can be established in a different way, such as reloading of specifications and argument selection of start processes.

Apart from that the simulator should be able to simulate PSF specifications (or rather a compiled form) according to the semantics, it must at least fulfill the following requirements.

- Simple interaction with the user for choosing an action to be executed from a list of possible executable actions at a certain moment. Simple in the way that the actions are presented in a single unordered list.

- Show on request the status of processes currently being simulated in a way that their correlation is visible and how the list of possible actions is determined from them.

- Make it possible to trace certain actions as they are executed. These actions must be selected from all actions in an easy manner.

- Be able to run randomly and stop this whenever one or more breakpoints are encountered. That can be on execution of an action on which a breakpoint is set, when one or more actions with breakpoints on them appear in the list of possible action, or when all actions in the list have breakpoints on them (synchronization). Selection of breakpoints should be made easy, preferably in a similar way of selecting actions to be traced.

- A history mechanism that not only makes it possible to undo or redo a step, but also to go to a previously marked state.

*3.2 Non-functional Requirements*

The non-functional requirements we list here represent our wishes as opposed to the implementation of the old simulator.

- A modular design with easy to replace components. Especially, the simulator should have a separate kernel which can be used in other applications.
- Can be used as a framework for simulating other languages similar to PSF, or variants of PSF.
- The user interface should be less dependent on the X Window System as is the case with the old simulator, and should be easy to adapt to changes in environment, application, user demands.
- Easy coupling of the simulator with animation.

**4. Architecture Specification of the Simulator**

We specify the architecture in several steps, starting with the architecture of a simple simulator to which we add the features. The architecture specification as presented here is the result of normal software development processes[1] incorporated with an architecture phase. In these processes there is feedback from following phases, and so also the architecture phase gets this feedback.

*4.1 A Simple Simulator*

Our simple simulator consists of four system components.

| | |
|---|---|
| kernel | does the actual simulation. |
| startprocess | takes care of choosing a process to start the simulating with. |
| actionchooser | takes care of choosing an action from a list of possible actions it receives from the kernel. |
| display | displays the information the other components wish to communicate to the user. |

We first specify the id's for the four components and the data, in an abstract form, that are used in the communication between them in a separate module.

```
data module SimulatorData
begin
   exports
   begin
      functions
         kernel : → ID
         startprocess : → ID
         actionchooser : → ID
         display : → ID

         start-process : → DATA
         action-choose-list : → DATA
         action : → DATA
         halt : → DATA
         reset : → DATA
   end
   imports
      ArchitectureTypes
end SimulatorData
```

The kernel can be in two states. One in which it actually simulates, and one in which it is waiting for communication with other components. This is specified by way of a boolean variable `wait`.

```
process module Kernel
begin
```

---

1. See [34] for an overview.

```
        exports
        begin
            processes
                Kernel
        end
        imports
            SimulatorData,
            ArchitecturePrimitives,
            Booleans
        atoms
            compute-choose-list
            compute-halt
        processes
            Kernel : BOOLEAN
        variables
            wait : → BOOLEAN
        definitions
            Kernel = Kernel(true)
            Kernel(wait) =
                (
                    [wait = false]→ (
                        compute-choose-list .
                        snd(kernel >> actionchooser, action-choose-list)
                    + compute-halt .
                        snd(kernel >> display, halt)
                    ) .
                    Kernel(true)
                + [wait = true]→ (
                        rec(actionchooser >> kernel, action) .
                        Kernel(false)
                    + rec(startprocess >> kernel, start-process) .
                        snd(kernel >> display, start-process) .
                        snd(kernel >> actionchooser, reset) .
                        Kernel(false)
                    )
                )
    end Kernel
```

If the kernel is not in the wait state, there is a choice between two internal actions. The action `compute-choose-list`, resembling the computing of a list of possible actions that can occur. This list is sent to the actionchooser. And the other action `compute-halt`, indicating the kernel could not compute a list of possible action, either because simulation ended, or a deadlock occurred. In the `wait` state it can receive a `start-process` from the startprocess component, or it can receive an `action` from the actionchooser.

The startprocess component is very simple, it can only send a `start-process` to the kernel.

```
    process module StartProcess
    begin
        exports
        begin
            processes
                StartProcess
        end
        imports
            ArchitecturePrimitives,
            SimulatorData
        atoms
            select-start-process
        definitions
            StartProcess =
                (
                    select-start-process .
                    snd(startprocess >> kernel, start-process)
                ) * delta
    end StartProcess
```

The actionchooser can receive an `action-choose-list` or a `reset` from the kernel. When it receives an `action-choose-list` it can send an `action` to the kernel.

```
    process module ActionChooser
    begin
```

```
exports
begin
    processes
        ActionChooser
end
imports
    ArchitecturePrimitives,
    SimulatorData,
    Booleans
atoms
    choose-action
processes
    Choose : BOOLEAN
    Reset
variables
    choose : → BOOLEAN
definitions
    ActionChooser = Choose(false)
    Choose(choose) =
            rec(kernel >> actionchooser, action-choose-list) .
            Choose(true)
        + [choose = true]→ (
            choose-action .
            (
                snd(actionchooser >> kernel, action) .
                Choose(false)
              + Reset
            )
          )
        + Reset
    Reset = rec(kernel >> actionchooser, reset) .
            Choose(false)
end ActionChooser
```

The possibility for a reset after an action has been chosen is necessary, otherwise a deadlock can occur when the kernel sends a reset caused by the receiving of a start-process.

The display can only receive from other components. At the moment it receives only from the kernel.

```
process module Display
begin
    exports
    begin
        processes
            Display
    end
    imports
        ArchitecturePrimitives,
        SimulatorData
    definitions
        Display =
            (
                rec(kernel >> display, halt)
              + rec(kernel >> display, start-process)
            ) * delta
end Display
```

We combine the components to a system by merging the processes of the components.

```
process module SimulatorSystem
begin
    exports
    begin
        processes
            SimulatorSystem
    end
    imports
        Kernel,
        StartProcess,
        ActionChooser,
        Display
    definitions
        SimulatorSystem =
```

```
                Kernel
             ‖  StartProcess
             ‖  ActionChooser
             ‖  Display
    end SimulatorSystem
```

We complete the architecture of the simple simulator by putting the system in the architecture environment.

```
    process module Simulator
    begin
        imports
            Architecture {
                System bound by [
                    System → SimulatorSystem
                ] to SimulatorSystem
                renamed by [
                    Architecture → Simulator
                ]
            }
    end Simulator
```

An animation of the architecture is shown in figure 2.

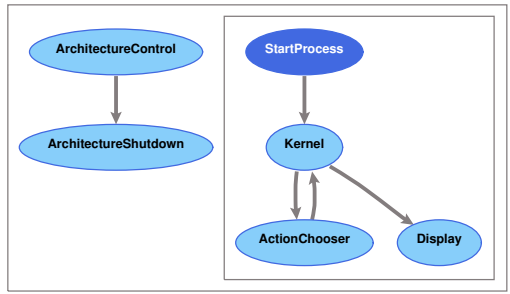

**Figure 2.** Architecture of a simple simulator

### 4.2 Functions

We extend the simple simulator with functions that can be invoked by the user, `quit` and `process-status`.

To module SimulatorData we add the id `function` and data terms for the functions. And we add a module Function.

```
    process module Function
    begin
        exports
        begin
            processes
                Function
        end
        imports
            ArchitecturePrimitives,
            SimulatorData
        atoms
            push-quit
            push-process-status
        definitions
            Function =
                (
                    push-quit .
                    snd(function >> kernel, quit)
                +   push-process-status .
                    snd(function >> kernel, process-status)
                ) * delta
    end Function
```

To module Kernel we add the following alternatives to the wait state.

```
              + rec(function >> kernel, quit) .
                snd-quit
              + rec(function >> kernel, process-status) .
                snd(kernel >> display, process-status) .
                Kernel(wait)
```

After the kernel receives a `quit` it communicates with the architecture environment by means of a `snd-quit` on which the environment acts with a shutdown. And on receiving `process-status` it send the process status to the display (we use the same abstract data term here).

To the module Display we add an alternative for receiving a `process-status` message and in the module SimulatorSystem we merge the process Function with the other processes. The animation of the resulting architecture is shown in figure 3.

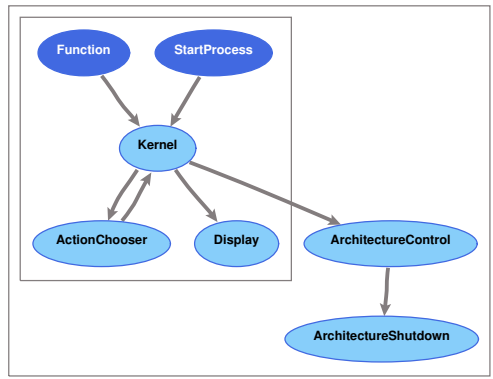

**Figure 3.** Architecture with functions

*4.3 Tracing*

We now add a component `tracectrl` that takes care of the tracing of actions (make them visible to the user) the moment they are executed. Whenever an action is chosen by the actionchooser it is send to tracectrl which decides, on indication by the user, whether it has to be traced, in which case a message is send to display. So it acts as a filter.

To module SimulatorData we add the id `tracectrl` and as data terms `trace-action` and `done`.

```
process module TraceCtrl
begin
   exports
   begin
      processes
         TraceCtrl
   end
   imports
      SimulatorData,
      ArchitecturePrimitives
   atoms
      trace
      no-trace
   definitions
      TraceCtrl =
         (
            rec(actionchooser >> tracectrl, action) .
            (
               trace .
               snd(tracectrl >> display, trace-action)
            + no-trace
            ) .
            snd(tracectrl >> actionchooser, done)
         ) * delta
end TraceCtrl
```

The confirmation to the actionchooser is necessary, otherwise it is possible the actionchooser continues and another message to the display is sent before a trace message is sent, and so a mix-up of the order of the messages on the display can occur.

We add the communication with tracectrl in the actionchooser directly after `action` is send to the kernel, as shown below with existing code in grey.

```
+ [choose = true]→ (
    snd(actionchooser >> kernel, action) .
    snd(actionchooser >> tracectrl, action) .
    rec(tracectrl >> actionchooser, done) .
    Choose(false)
)
```

To module Display we add an alternative for receiving a `trace-action` message and we add TraceCtrl to SimulatorSystem. The resulting architecture is shown in figure 4.

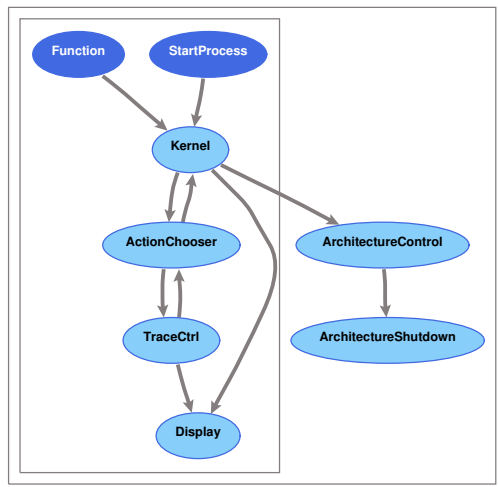

**Figure 4.** Architecture with tracing

*4.4 Random*

At this moment it is of no concern whether the user wants to let the actionchooser choose actions randomly, so this can be kept implicit with the actionchooser. But when we introduce breakpoints in order to stop the simulator from running randomly at certain moments, we need to know whether the simulator is running randomly explicitly. So we add a control state to the Choose process of the actionchooser and the possibility to switch random on and off.

```
process module ActionChooser
begin
    exports
    begin
        processes
            ActionChooser
    end
    imports
        ArchitecturePrimitives,
        SimulatorData,
        Booleans
    atoms
        choose-action
        random-on
        random-off
    processes
        Choose : BOOLEAN # BOOLEAN
        Reset : BOOLEAN
```

```
       variables
           random :  → BOOLEAN
           choose :  → BOOLEAN
       definitions
           ActionChooser = Choose(false, false)
           Choose(random, choose) =
                 rec(kernel >> actionchooser, action-choose-list) .
                 Choose(random, true)
              +  [choose = true]→ (
                    choose-action .
                    (
                       snd(actionchooser >> kernel, action) .
                       snd(actionchooser >> tracectrl, action) .
                       rec(tracectrl >> actionchooser, done) .
                       Choose(random, false)
                    +  Reset(random)
                    )
                 )
              +  Reset(random)
              +  [random = true]→ (
                    random-off .
                    Choose(false, choose)
                 )
              +  [random = false]→ (
                    random-on .
                    Choose(true, choose)
                 )
           Reset(random) =
                 rec(kernel >> actionchooser, reset) .
                 Choose(random, false)
     end ActionChooser
```

## 4.5 Breakpoints

In order to stop the simulator from running randomly at certain moments we add breakpoints. There are two type of breakpoints. One is when an action (indicated by the user) gets executed, and the other is when the list of possible actions contains one or more actions on which the user has set a breakpoint.

To module SimulatorData we add the id `breakctrl` and as data terms `break-action`, `break` end `no-break`.

```
       process module BreakCtrl
       begin
          exports
          begin
             processes
                BreakCtrl
          end
          imports
             SimulatorData,
             ArchitecturePrimitives
          atoms
             break
             no-break
             break-list
             no-break-list
          definitions
             BreakCtrl =
                (
                   rec(actionchooser >> breakctrl, action) .
                   (
                      break .
                      snd(breakctrl >> display, break-action) .
                      snd(breakctrl >> actionchooser, break)
                   +  no-break .
                      snd(breakctrl >> actionchooser, no-break)
                   )
                +  rec(actionchooser >> breakctrl, action-choose-list) .
                   (
                      no-break-list .
```

```
                    snd(breakctrl >> actionchooser, action-choose-list)
                 +  break-list .
                    snd(breakctrl >> display, break) .
                    snd(breakctrl >> actionchooser, break)
                 )
              ) * delta
       end BreakCtrl
```

In module ActionChooser we replace

```
              rec(kernel >> actionchooser, action-choose-list) .
              Choose(random, true)
```

with

```
              rec(kernel >> actionchooser, action-choose-list) .
              (
                 [random = true]→ (
                    snd(actionchooser >> breakctrl, action-choose-list) .
                    (
                       rec(breakctrl >> actionchooser, break) .
                       force-random-off .
                       present-list .
                       Choose(false, true)
                    +  rec(breakctrl >> actionchooser, action-choose-list) .
                       present-list .
                       Choose(true, true)
                    )
                 )
              +  [random = false]→ (
                    present-list .
                    Choose(false, true)
                 )
              )
```

and

```
              snd(actionchooser >> tracectrl, action) .
              rec(tracectrl >> actionchooser, done) .
              Choose(random, false)
```

with

```
              (
                 [random = true]→ (
                    snd(actionchooser >> breakctrl, action) .
                    (
                       rec(breakctrl >> actionchooser, break) .
                       force-random-off .
                       Choose(false, false)
                    +  rec(breakctrl >> actionchooser, no-break) .
                       snd(actionchooser >> tracectrl, action) .
                       rec(tracectrl >> actionchooser, done) .
                       Choose(true, false)
                    )
                 )
              +  [random = false]→ (
                    snd(actionchooser >> tracectrl, action) .
                    rec(tracectrl >> actionchooser, done) .
                    Choose(false, false)
                 )
              )
```

We also add the introduced actions `forced-random-off` and `present-list` to the atoms section of the module ActionChooser. The action `force-random-off` is necessary because it clearly differs from `random-off` which is invoked by the user. The action `present-list` has a more complex explanation. In the old situation this action could be combined with the receiving of the `action-choose-list`, we now have to do later in the process. This becomes more clear when we are going to refine the actions in a later stage (see section 7.1).

To module Display we add alternatives for receiving a `break-action` and a `break` message and we add BreakCtrl to SimulatorSystem. The resulting architecture is shown in figure 5.

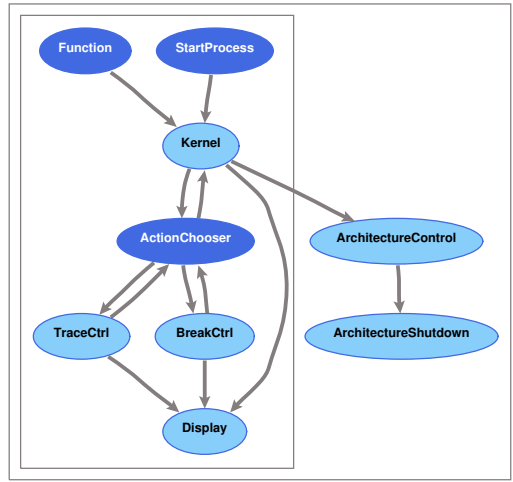

**Figure 5.** Architecture with breakpoints

## 5. ToolBus Application Design with PSF

We give in this section the specification of the PSF ToolBus library which appeared earlier in [15] followed by a specification of our toy example as ToolBus application for which we specified the architecture in section 2.2.

### 5.1 Specification of the PSF ToolBus library

This section presents a specification of a library of interfaces for PSF which can be used as a basis for the specification of ToolBus applications. This specification does not cover all the facilities of the ToolBus, but just what is necessary for the project at hand.

#### 5.1.1 Data

First, a sort is defined for the data terms used in the tools. An abstraction is made from the actual data used by the tools.

```
data module ToolTypes
begin
   exports
   begin
      sorts
         Tterm
   end
end ToolTypes
```

Next, the sorts are introduced for the data terms and identifiers which will be used inside the ToolBus as well as for communication with the ToolBus.

```
data module ToolBusTypes
begin
   exports
   begin
      sorts
         TBterm,
         TBid
   end
end ToolBusTypes
```

The module ToolFunctions provides names for conversions between data terms used outside and inside the ToolBus.

```
data module ToolFunctions
begin
   exports
   begin
      functions
         tbterm : Tterm → TBterm
         tterm : TBterm → Tterm
   end
   imports
      ToolTypes,
      ToolBusTypes
   variables
      t : → Tterm
   equations
   ['] tterm(tbterm(t)) = t
end ToolFunctions
```

The ToolBus has access to several functions operating on different types. Here only the operators for tests on equality and inequality of terms, will be needed. These are introduced in the module ToolBusFunctions.

```
data module ToolBusFunctions
begin
   exports
   begin
      functions
         equal : TBterm # TBterm → BOOLEAN
   end
   imports
      ToolBusTypes,
      Booleans
   variables
      tb1 : → TBterm
      tb2 : → TBterm
   equations
   ['] equal(tb1, tb1) = true
   ['] not(equal(tb1, tb2)) = true
end ToolBusFunctions
```

*5.1.2 Connecting tools to the ToolBus*

In figure 6 two possible ways of connecting tools to the ToolBus are displayed. One way is to use a separate adapter and the other to have a builtin adapter. Tool1 communicates with its adapter over pipelines.[2]

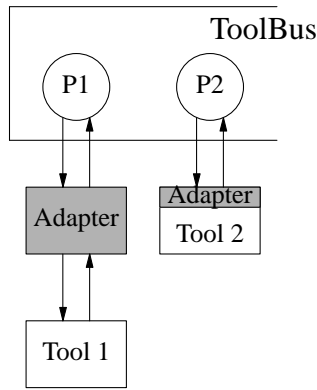

**Figure 6.** Model of tool and ToolBus interconnection

---

2. In Unix systems, a pipeline is a means of communication between two processes.

Next we define the primitives for communication between a tool and its adapter.

```
process module ToolAdapterPrimitives
begin
    exports
    begin
        atoms
            tooladapter-rec : Tterm
            tooladapter-snd : Tterm
    end
    imports
        ToolTypes
end ToolAdapterPrimitives
```

The primitives for communication between a tool and the ToolBus are fixed by the ToolBus design. At this stage these need to be formally defined in PSF, however. These primitives can be used for communication between an adapter and the ToolBus as well, since the adapter logically takes the place of the tool it is supposed to connect to the ToolBus.

```
process module ToolToolBusPrimitives
begin
    exports
    begin
        atoms
            tooltb-snd : TBterm
            tooltb-rec : TBterm

            tooltb-snd-event : TBterm
            tooltb-rec-ack-event : TBterm
    end
    imports
        ToolBusTypes
end ToolToolBusPrimitives
```

Inside a ToolBus script a number of primitives may be used consisting of the actions for communication between ToolBus processes and their synchronous communication action, the actions used to communicate with the tools, and the action required to shutdown the ToolBus.

```
process module ToolBusPrimitives
begin
    exports
    begin
        atoms
            tb-snd-msg : TBterm # TBterm
            tb-rec-msg : TBterm # TBterm
            tb-comm-msg : TBterm # TBterm
            tb-snd-msg : TBterm # TBterm # TBterm
            tb-rec-msg : TBterm # TBterm # TBterm
            tb-comm-msg : TBterm # TBterm # TBterm

            tb-snd-eval : TBid # TBterm
            tb-rec-value : TBid # TBterm
            tb-snd-do : TBid # TBterm
            tb-rec-event : TBid # TBterm
            tb-snd-ack-event : TBid # TBterm

            tb-shutdown
    end
    imports
        ToolBusTypes
    communications
        tb-snd-msg(tb1, tb2) | tb-rec-msg(tb1, tb2) = tb-comm-msg(tb1, tb2)
            for tb1 in TBterm, tb2 in TBterm
        tb-snd-msg(tb1, tb2, tb3) | tb-rec-msg(tb1, tb2, tb3) =
            tb-comm-msg(tb1, tb2, tb3)
            for tb1 in TBterm, tb2 in TBterm, tb3 in TBterm
end ToolBusPrimitives
```

The ToolBus provides primitives allowing an arbitrary number of terms as parameters for communication between processes in the ToolBus. Here, the specification only covers the case of two and three term arguments for the primitives, because versions with more are usually not needed. In order to do better lists of terms have to be introduced, which is entirely possible in PSF but an unnecessary complication at this stage. The two-term version can be used with the first term as a 'to' or 'from' identifier and the second as a data argument. The three-term version can be used with the first term as 'from', the second as 'to', and the third as the actual data argument. If more arguments have to be passed, they can always be grouped into a single argument.

The module NewTool is a generic module with parameter Tool for connecting a tool to the ToolBus.

```
process module NewTool
begin
    parameters
        Tool
        begin
            processes
                Tool
        end Tool
    exports
    begin
        atoms
            tooltb-snd-value : TBid # TBterm
            tooltb-rec-eval : TBid # TBterm
            tooltb-rec-do : TBid # TBterm
            tooltb-snd-event : TBid # TBterm
            tooltb-rec-ack-event : TBid # TBterm
        processes
            TBProcess
        sets
            of atoms
                TBProcess = {
                    tb-rec-value(tid, tb), tooltb-snd(tb),
                    tb-snd-eval(tid, tb), tb-snd-do(tid, tb),
                    tooltb-rec(tb), tb-rec-event(tid, tb),
                    tooltb-snd-event(tb), tb-snd-ack-event(tid, tb),
                    tooltb-rec-ack-event(tb)
                    | tid in TBid, tb in TBterm
                }
    end
    imports
        ToolToolBusPrimitives,
        ToolBusPrimitives
    communications
        tooltb-snd(tb) | tb-rec-value(tid, tb) = tooltb-snd-value(tid, tb)
            for t in TBterm, tid in TBid
        tooltb-rec(tb) | tb-snd-eval(tid, tb) = tooltb-rec-eval(tid, tb)
            for t in TBterm, tid in TBid
        tooltb-rec(tb) | tb-snd-do(tid, tb) = tooltb-rec-do(tid, tb)
            for t in TBterm, tid in TBid
        tooltb-snd-event(tb) | tb-rec-event(tid, tb) = tooltb-snd-event(tid, tb)
            for t in TBterm, tid in TBid
        tooltb-rec-ack-event(tb) | tb-snd-ack-event(tid, tb) =
            tooltb-rec-ack-event(tid, tb) for tb in TBterm, tid in TBid
    definitions
        TBProcess = encaps(TBProcess, Tool)
end NewTool
```

The process Tool accomplishes the connection between a process inside the ToolBus and a tool outside the ToolBus. The process TBProcess encapsulates the process Tool in order to enforce communications and thereby to prevent communications with other tools or processes. Note that TBProcess is used as the name of the main process and as the name of the encapsulation set. By doing so, they can both be renamed with a single renaming. This renaming is necessary if more than one tool is connected to the ToolBus (which is of course the whole point of the ToolBus).

The module NewToolAdapter is a generic module with parameters Tool and Adapter for connecting a tool and its adapter.

```
process module NewToolAdapter
begin
    parameters
        Tool
```

```
            begin
                atoms
                    tool-snd : Tterm
                    tool-rec : Tterm
                processes
                    Tool
            end Tool,
            Adapter
            begin
                processes
                    Adapter
            end Adapter
        exports
        begin
            atoms
                tooladapter-comm : Tterm
                adaptertool-comm : Tterm
            processes
                ToolAdapter
            sets
                of atoms
                    ToolAdapter = {
                       tool-snd(t), tooladapter-rec(t),
                       tool-rec(t), tooladapter-snd(t)
                        | t in Tterm
                    }
        end
        imports
            ToolAdapterPrimitives,
            ToolBusTypes
        communications
            tool-snd(t)  |  tooladapter-rec(t) = tooladapter-comm(t) for t in Tterm
            tool-rec(t)  |  tooladapter-snd(t) = adaptertool-comm(t) for t in Tterm
        definitions
            ToolAdapter = encaps(ToolAdapter, Adapter ‖ Tool)
    end NewToolAdapter
```

The process ToolAdapter puts an Adapter and a Tool in parallel and enforces communication between them by an encapsulation. In this case the main process and the encapsulation set have the same name once more, so that only one renaming is needed.

### 5.1.3 ToolBus instantiation

The module NewToolBus is a generic module with parameter Application for instantiation of the ToolBus with an application.

```
        process module NewToolBus
        begin
            parameters
                Application
                begin
                    processes
                        Application
                end Application
            exports
            begin
                processes
                    ToolBus
            end
            imports
                ToolBusPrimitives
            atoms
                application-shutdown
                tbc-shutdown
                tbc-app-shutdown
                TB-shutdown
                TB-app-shutdown
            processes
                ToolBus-Control
                Shutdown
```

```
            sets
                of atoms
                    H = {
                        tb-snd-msg(tb1, tb2), tb-rec-msg(tb1, tb2),
                        tb-snd-msg(tb1, tb2, tb3), tb-rec-msg(tb1, tb2, tb3)
                        | tb1 in TBterm, tb2 in TBterm, tb3 in TBterm
                    }
                    TB-H = {
                        tb-shutdown, tbc-shutdown,
                        tbc-app-shutdown, application-shutdown
                    }
                    P = { TB-shutdown, TB-app-shutdown }
            communications
                tb-shutdown | tbc-shutdown = TB-shutdown
                tbc-app-shutdown | application-shutdown = TB-app-shutdown
            definitions
                ToolBus =
                    encaps(TB-H,
                        prio(P > atoms,
                            ToolBus-Control
                        || disrupt(
                                encaps(H, Application),
                                Shutdown
                            )
                        )
                    )
                ToolBus-Control = tbc-shutdown . tbc-app-shutdown
                Shutdown = application-shutdown
        end NewToolBus
```

A toolbus application can be described more clearly with `ToolBus = encaps(H, Application)`. The remaining code is needed to force a shutdown of all processes that otherwise would be left either running or in a state of deadlock after a ToolBus shutdown by the application. When an application needs to shutdown it performs an action `tb-shutdown` which will communicate with the action `tbc-shutdown` of the `ToolBus-Control` process, which then performs a `tbc-app-shutdown` that will communicate with `application-shutdown` of the `Shutdown` process enforcing a disrupt of the `Application` process.

In figure 7 an overview is given of the import relations of the modules in the PSF ToolBus library. The module Booleans stems from a standard library of PSF.

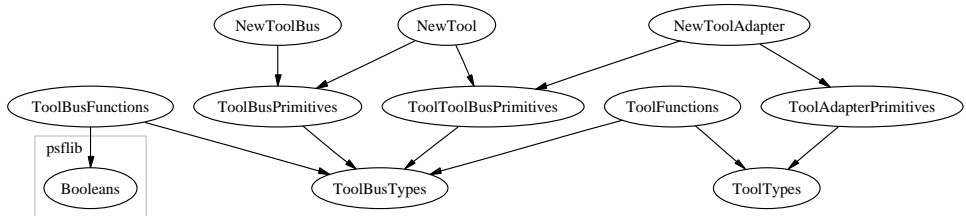

**Figure 7.** Import graph of the ToolBus library

*5.2 Example*

As an example of the use of the PSF ToolBus library, the specification is given of an application of which we specified the architecture in section 2.2, carried out in the form as shown in figure 6. In this example, Tool1 can either send a 'message' to Tool2 and then wait for an acknowledgement from Tool2, or it can send a 'quit' after which the application will shutdown.

*5.2.1 Specification of the tools*

The first module defines the data that will be used.

```
data module Data
begin
    exports
    begin
        functions
            message : → Tterm
            ack : → Tterm
            quit : → Tterm
    end
    imports
        ToolTypes
end Data
```

A specification of Tool1 and its adapter is then obtained.

```
process module Tool1
begin
    exports
    begin
        atoms
            snd : Tterm
            rec : Tterm
        processes
            Tool1
    end
    imports
        Data
    definitions
        Tool1 =
            snd(message) .
            rec(ack) .
            Tool1
        +   snd(quit)
end Tool1

process module AdapterTool1
begin
    exports
    begin
        processes
            AdapterTool1
    end
    imports
        Data,
        ToolFunctions,
        ToolAdapterPrimitives,
        ToolToolBusPrimitives
    definitions
        AdapterTool1 =
            tooladapter-rec(message) .
            tooltb-snd-event(tbterm(message)) .
            tooltb-rec-ack-event(tbterm(message)) .
            tooladapter-snd(ack) .
            AdapterTool1
        +   tooladapter-rec(quit) .
            tooltb-snd-event(tbterm(quit))
end AdapterTool1
```

Tool1 and its adapter are combined by importing NewToolAdapter and binding the parameters.

```
process module Tool1Adapter
begin
    imports
        NewToolAdapter {
            Tool bound by [
                tool-snd → snd,
                tool-rec → rec,
                Tool → Tool1
            ] to Tool1
            Adapter bound by [
                Adapter → AdapterTool1
            ] to AdapterTool1
            renamed by [
```

```
                    ToolAdapter → Tool1Adapter
                ]
        }
    end Tool1Adapter
```

We specify Tool2

```
    process module Tool2
    begin
        exports
        begin
            processes
                Tool2
        end
        imports
            Data,
            ToolFunctions,
            ToolToolBusPrimitives
        definitions
            Tool2 =
                tooltb-rec(tbterm(message)) .
                tooltb-snd(tbterm(ack)) .
                Tool2
    end Tool2
```

*5.2.2 Specification of the ToolBus processes*

Some identifiers are defined in order to distinguish the messages sent between ToolBus processes themselves and between ToolBus processes and their accompanying tools. The lowercase identifiers (of type TBterm) are used with the actions `tb-snd-msg` and `tb-rec-msg`. The first argument of a message will always be the origin of the message, and the second argument will serve as its destination. Uppercase identifiers (of type TBid) are used as tool identifiers. Strictly speaking these are not necessary, since there can't be any communication with any other tool because of encapsulation. By using them, however, the actions for communication with a tool will have more similarity to the ones used in the ToolBus.

```
    data module ID
    begin
        exports
        begin
            functions
                T1 :  → TBid
                t1 :  → TBterm
                T2 :  → TBid
                t2 :  → TBterm
        end
        imports
            ToolBusTypes
    end ID
```

For both tools a ToolBus process is defined. The specifications for these processes describe the protocol for communication between the tools.

```
    process module PTool1
    begin
        exports
        begin
            processes
                PTool1
        end
        imports
            Tool1Adapter,
            ID,
            ToolBusPrimitives,
            ToolBusFunctions
        processes
            PT1
        definitions
            PTool1 = Tool1Adapter ∥ PT1
```

```
            PT1 =
                    tb-rec-event(T1, tbterm(message)) .
                    tb-snd-msg(t1, t2, tbterm(message)) .
                    tb-rec-msg(t2, t1, tbterm(ack)) .
                    tb-snd-ack-event(T1, tbterm(message)) .
                    PT1
                + tb-rec-event(T1, tbterm(quit)) .
                    snd-tb-shutdown
    end PTool1

    process module PTool2
    begin
        exports
        begin
            processes
                PTool2
        end
        imports
            Tool2,
            ID,
            ToolBusPrimitives
        processes
            PT2
        definitions
            PTool2 = Tool2 ‖ PT2
            PT2 =
                tb-rec-msg(t1, t2, tbterm(message)) .
                tb-snd-eval(T2, tbterm(message)) .
                tb-rec-value(T2, tbterm(ack)) .
                tb-snd-msg(t2, t1, tbterm(ack)) .
                PT2
    end PTool2
```

*5.2.3 Specification of the ToolBus application*

The ToolBus processes are connected with the tools and together they constitute the process System that merges the resulting two processes.

```
    process module Tools
    begin
        exports
        begin
            processes
                System
        end
        imports
            NewTool {
                Tool bound by [
                    Tool → PTool1
                ] to PTool1
                renamed by [
                    TBProcess → XPTool1
                ]
            },
            NewTool {
                Tool bound by [
                    Tool → PTool2
                ] to PTool2
                renamed by [
                    TBProcess → XPTool2
                ]
            },
            ID,
            ToolBusFunctions
        definitions
            System = XPTool1 ‖ XPTool2
    end Tools
```

At this stage renamings are necessary to be able to distinguish the two processes TBProcess (and sets).

The process System is now transformed into a ToolBus application.

```
process module App
begin
   imports
      NewToolBus {
         Application bound by [
            Application → System
         ] to Tools
      }
end App
```

The main process of this application is ToolBus. A generated animation is shown in figure 8, in which AdapterTool1 just sent a message it had received from Tool1, to ToolBus process PT1.

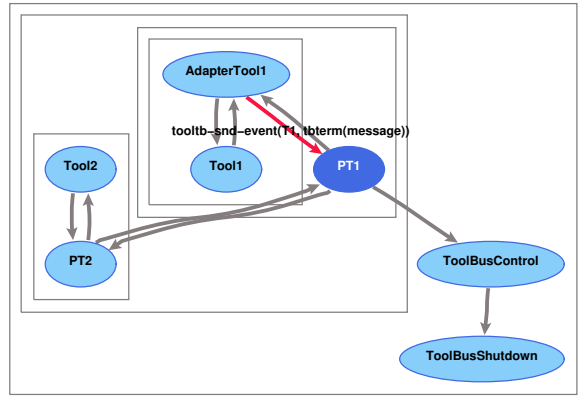

**Figure 8.** Animation of the ToolBus specification example

*5.2.4 Example as ToolBus application*

The application we have specified above has been implemented as an application consisting of three Tcl/Tk [30] programs (Tool1, its adapter, and Tool2), and a ToolBus script. A screendump of this application at work together with the viewer[3] of the ToolBus is shown in figure 9. The ToolBus script is shown below. The processes `PT1` and `PT2` closely resemble the processes `PTool1` and `PTool2` in our PSF specification. The `execute` actions in the ToolBus script correspond to starting of the adapter for Tool1 and starting of Tool2 in parallel with the processes `PT1` and `PT2` respectively.

```
process PT1 is
let
   T1: tool1adapter
in
   execute(tool1adapter, T1?) .
   (
      rec-event(T1, message) .
      snd-msg(t1, t2, message) .
      rec-msg(t2, t1, ack) .
      snd-ack-event(T1, message)
   +  rec-event(T1, quit) .
      shutdown("")
   ) * delta
endlet

process PT2 is
let
   T2: tool2
in
   execute(tool2, T2?) .
   (
```

---

3. With the viewer it is possible to step through the execution of the ToolBus script and view the variables of the individual processes inside the ToolBus.

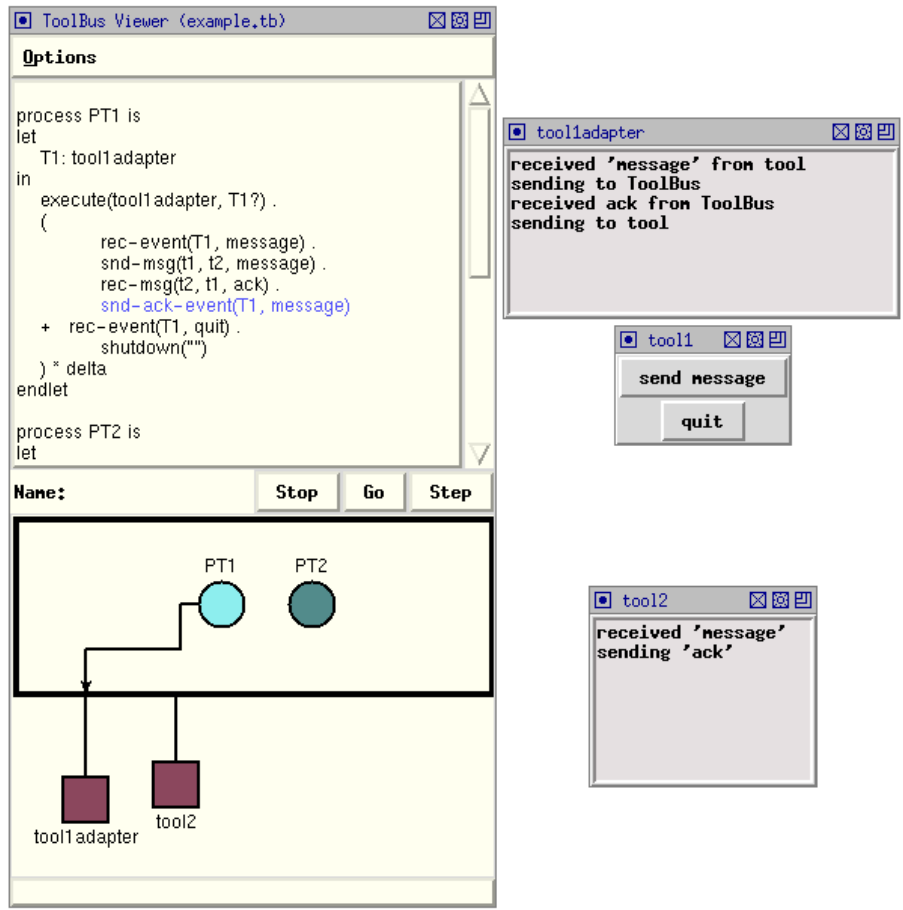

**Figure 9.** Screendump of the example as ToolBus application with viewer

```
            rec-msg(t1, t2, message) .
            snd-eval(T2, eval(message)) .
            rec-value(T2, value(ack)) .
            snd-msg(t2, t1, ack)
      ) * delta
endlet

tool tool1adapter is { command = "wish-adapter -script tool1adapter.tcl" }
tool tool2 is { command = "wish-adapter -script tool2.tcl" }

toolbus(PT1, PT2)
```

The actions `snd-eval` and `rec-value` differentiate from their equivalents in the PSF specification. The term `eval(message)` instead of just `message` is needed because the interpreter of evaluation requests that a tool receives from the ToolBus, calls a function with the name it finds as function in this term. We could have used any name instead of `eval` provided that Tool2 has got a function with that name.
Why the same scheme is needed by the ToolBus for `rec-value` is not known.

The processes in the ToolBus script use iteration and the processes in the PSF specification recursion. In PSF it is also possible to use iteration in this case, since the processes have no arguments to hold the current state. On the other hand, in PSF it is not possible to define variables for storing a global state, so when it is necessary to hold the current state, this must be done through the arguments of a process and be formalized via recursion.

The last line of the ToolBus script starts the processes `PT1` and `PT2` in parallel. Its equivalent in the PSF

specification is the process `System`.

## 6. From Architecture to ToolBus Application Design

It is only useful to invest a lot of effort in the architecture if we can relate it to a design on a lower level. In this section we describe the techniques we use to come from an architecture specification to a ToolBus application specification. We demonstrate these techniques with our toy example.

### 6.1 Horizontal Implementation

Given two processes $S$ and $I$, $I$ is an implementation of $S$ if $I$ is more deterministic (or equivalent). As the actions $S$ and $I$ perform belong to the same alphabet, $S$ and $I$ belong to the same abstraction level. Such an implementation relation is called *horizontal*.

To achieve a horizontal implementation we use parallel composition, which can be used to constrain a process. Consider process $P = a \cdot P$, which can do action $a$ at every moment. If we put $P$ in parallel with the process $Q = x \cdot b \cdot Q$ with communication $a \mid b = c$ and enforcing the communication by encapsulation, process $P$ can only do action $a$ whenever process $Q$ has first done action $x$. So process $P$ is constrained by $Q$ and $P \parallel Q$ is an horizontal implementation of $P$, provided $Q$ only interacts with $P$ through $b$. This form of controlling a process is also known as *superimposition* [9] or *superposition* [19] as composition.

### 6.2 Vertical Implementation

In [31], action refinement is used as a technique for mapping abstract actions onto concrete processes, called *vertical* implementation, which is more fully described in [32]. With vertical implementation we want to relate processes that belong to different abstraction levels, where the change of level usually comes with a change of alphabet. For such processes we like to develop *vertical* implementation relations that, given an abstract process $S$ and a concrete process $I$, tells us if $I$ is an implementation for the specification $S$. More specifically, we want to develop a mapping of abstract actions to sequences of one or more concrete actions so that $S$ and $I$ are *vertical bisimular*.

We give a rationale of vertical implementation. Consider the processes $P = a \cdot b$ with $a$ an internal action and $Q = c \cdot d \cdot e$ with internal actions $c$, and $d$. If we refine abstract action $a$ from process $P$ to the sequence of concrete actions $c \cdot d$ and rename action $b$ to $e$ we obtain process $Q$. We consider the processes $P$ and $Q$ vertical bisimular with respect to the mapping consisting of the above refinement and renaming.

We can explain the notion of *vertical bisimular* by the following. We hide the internal action $a$ of process $P$ by replacing it with the silent step $\tau$ to obtain $P = \tau \cdot b$. Applying the algebraic law $x \cdot \tau = x$ gives us $P = \tau \cdot \tau \cdot b$. If we now replace the first $\tau$ with $c$ and the second with $d$, and rename $b$ into $e$ we obtain the process $Q$. With $H$ as hide operator and $R$ as renaming operator we can prove that $R_{\{b \to e\}}(H_{\{a\}}(P))$ and $H_{\{c,d\}}(Q)$ are rooted weak bisimular. So vertical bisimulation is built on rooted weak bisimulation as horizontal implementation relation.

### 6.3 Example

Take the process `Component1` from the architecture of our toy example.

```
Component1 =
    send-message .
    snd(c1 >> c2, message) .
    rec(c2 >> c1, ack) .
    Component1
  + stop .
    snd-quit
```

We can make a virtual implementation by applying the following mapping.

```
send-message          → tb-rec-event(T1, tbterm(message))
```

```
snd(c1 >> c2, message) → tb-snd-msg(t1, t2, tbterm(message))
rec(c2 >> c1, ack)     → tb-rec-msg(t2, t1, tbterm(ack)) .
                         tb-snd-ack-event(T1, tbterm(message))
stop                   → tb-rec-event(T1, tbterm(quit))
snd-quit               → snd-tb-shutdown
```

And renaming `Component1` into `PT1` gives the following result.

```
PT1 =
    tb-rec-event(T1, tbterm(message)) .
    tb-snd-msg(t1, t2, tbterm(message)) .
    tb-rec-msg(t2, t1, tbterm(ack)) .
    tb-snd-ack-event(T1, tbterm(message)) .
    PT1
 +  tb-rec-event(T1, tbterm(quit)) .
    snd-tb-shutdown
```

We now make a horizontal implementation by constraining `PT1` with `Tool1Adapter`.

```
PTool1 = Tool1Adapter ∥ PT1
```

An implementation for `Component2` can be obtained in a similar way.

## 7. System Specification of the Simulator

We take the specification of the architecture of the simulator and turn it into a specification of a ToolBus application with the use of the techniques described in the previous chapter.

### 7.1 Refining

We show here the mapping for the virtual implementation of the architecture specification. We start with some default mappings that only apply when there are no other mappings to apply.

```
snd($1 >> $2, $3)  → tb-snd-msg($1, $2, tbterm($3))
rec($1 >> $2, $3)  → tb-rec-msg($1, $2, tbterm($3))
```

The `$n` on the left hand side represent matched terms that have to be filled in on the right hand side. Below the mappings per module are given.

module Kernel

```
compute-choose-list    →
                       tb-snd-eval(KERNEL, tbterm(compute-choose-list))
action-choose-list → tb-rec-value(KERNEL, tbterm(action-choose-list))
halt                 → tb-rec-value(KERNEL, tbterm(halt))
rec(actionchooser >> kernel, action)   →
                       tb-rec-msg(actionchooser, kernel, tbterm(action)) .
                       tb-snd-do(KERNEL, tbterm(action))
rec(function >> kernel, quit) →
                       tb-rec-msg(function, kernel, tbterm(quit)) .
                       tb-snd-do(KERNEL, tbterm(quit))
snd-quit             → snd-tb-shutdown
rec(function >> kernel, process-status)  →
                       tb-rec-msg(function, kernel, tbterm(process-status)) .
                       tb-snd-eval(KERNEL, tbterm(process-status)) .
                       tb-rec-value(KERNEL, tbterm(process-status))
rec(startprocess >> kernel, start-process)→
                       tb-rec-msg(startprocess, kernel,
                          tbterm(start-process)) .
                       tb-snd-do(KERNEL, tbterm(start-process))
```

module StartProcess

```
select-start-process →
                       tb-rec-event(STARTPROCESS, tbterm(start-process)) .
                       tb-snd-ack-event(STARTPROCESS, tbterm(start-process))
```

module ActionChooser

```
force-random-off   → tb-snd-do(ACTIONCHOOSER, tbterm(random-off))
```

```
present-list        → tb-snd-do(ACTIONCHOOSER, tbterm(action-choose-list))
choose-action       → tb-rec-event(ACTIONCHOOSER, tbterm(action)) .
                      tb-snd-ack-event(ACTIONCHOOSER, tbterm(action))
rec(kernel >> actionchooser, reset) →
                      tb-rec-msg(kernel, actionchooser, tbterm(reset)) .
                      tb-snd-do(ACTIONCHOOSER, tbterm(reset))
random-off          → tb-rec-event(ACTIONCHOOSER, tbterm(random-off)) .
                      tb-snd-ack-event(ACTIONCHOOSER, tbterm(random-off))
random-on           → tb-rec-event(ACTIONCHOOSER, tbterm(random-on)) .
                      tb-snd-ack-event(ACTIONCHOOSER, tbterm(random-on))
```

module Function

```
push-quit           → tb-rec-event(FUNCTION, tbterm(quit)) .
                      tb-snd-ack-event(FUNCTION, tbterm(quit))
push-process-status →
                      tb-rec-event(FUNCTION, tbterm(process-status)) .
                      tb-snd-ack-event(FUNCTION, tbterm(process-status))
```

module TraceCtrl

```
rec(actionchooser >> tracectrl, action)  →
                      tb-rec-msg(actionchooser, tracectrl, tbterm(action)) .
                      tb-snd-eval(TRACECTRL, tbterm(action))
trace               → tb-rec-value(TRACECTRL, tbterm(trace))
no-trace            → tb-rec-value(TRACECTRL, tbterm(no-trace))
```

module BreakCtrl

```
rec(actionchooser >> breakctrl, action)  →
                      tb-rec-msg(actionchooser, breakctrl, tbterm(action)) .
                      tb-snd-eval(BREAKCTRL, tbterm(action))
break               → tb-rec-value(BREAKCTRL, tbterm(break))
no-break            → tb-rec-value(BREAKCTRL, tbterm(no-break))
rec(actionchooser >> breakctrl, action-choose-list)→
                      tb-rec-msg(actionchooser, breakctrl,
                         tbterm(action-choose-list)) .
                      tb-snd-eval(BREAKCTRL, tbterm(action-choose-list))
break-list          → tb-rec-value(BREAKCTRL, tbterm(break))
no-break-list       → tb-rec-value(BREAKCTRL, tbterm(action-choose-list))
```

module Display

```
rec($1 >> display, $2)→
                      tb-rec-msg($1, display, tbterm($2)) .
                      tb-snd-do(DISPLAY, tbterm($2))
```

We rename all component modules and their main processes by putting a P in front of the name, indicating a Process in the ToolBus, to distinguish them from the tools for which we use a T in front of the name and possible adapters for which we use an A.

*7.2 Constraining*

We constrain the ToolBus processes obtained in the previous section with the specification of the tools. We confine ourselves to the constraining of the process PKernel, since the constraining of the other processes is rather straightforward and later we shall refine the Kernel even further. We show the module for the Kernel below. Here the main process PT-Kernel is the parallel composition of PKernel with the constraining process TKernel.

**process module** PKernel
**begin**
   **exports**
   **begin**
      **processes**
         PT-Kernel
   **end**
   **imports**
      SimulatorData,
      ToolBusPrimitives,
      ToolFunctions,
      TKernel,

```
        Booleans
    processes
        PKernel
        Kernel : BOOLEAN
    variables
        wait : → BOOLEAN
    definitions
        PT-Kernel = PKernel ∥ TKernel
        PKernel = Kernel(true)
        Kernel(wait) =
            (
                [wait = false]→ (
                    tb-snd-eval(KERNEL, tbterm(compute-choose-list)) .
                    (
                        tb-rec-value(KERNEL, tbterm(action-choose-list)) .
                        tb-snd-msg(kernel, actionchooser, tbterm(action-choose-list))
                      + tb-rec-value(KERNEL, tbterm(halt)) .
                        tb-snd-msg(kernel, display, tbterm(halt))
                    )
                ) .
                Kernel(true)
              + [wait = true]→ (
                    tb-rec-msg(actionchooser, kernel, tbterm(action)) .
                    tb-snd-do(KERNEL, tbterm(action)) .
                    Kernel(false)
                  + tb-rec-msg(function, kernel, tbterm(quit)) .
                    tb-snd-do(KERNEL, tbterm(quit)) .
                    snd-tb-shutdown
                  + tb-rec-msg(function, kernel, tbterm(process-status)) .
                    tb-snd-eval(KERNEL, tbterm(process-status)) .
                    tb-rec-value(KERNEL, tbterm(process-status)) .
                    tb-snd-msg(kernel, display, tbterm(process-status)) .
                    Kernel(true)
                  + tb-rec-msg(startprocess, kernel, tbterm(start-process)) .
                    tb-snd-do(KERNEL, tbterm(start-process)) .
                    tb-snd-msg(kernel, display, tbterm(start-process)) .
                    tb-snd-msg(kernel, actionchooser, tbterm(reset)) .
                    Kernel(false)
                )
            )
    end PKernel
```

Were the tools TKernel is specified as follows.

```
        process module TKernel
        begin
            exports
            begin
                processes
                    TKernel
            end
            imports
                SimulatorData,
                ToolToolBusPrimitives,
                ToolFunctions
            atoms
                action-choose-list
                halt
            definitions
                TKernel =
                    tooltb-rec(tbterm(compute-choose-list)) .
                    (
                        action-choose-list .
                        tooltb-snd(tbterm(action-choose-list))
                      + halt .
                        tooltb-snd(tbterm(halt))
                    ) . TKernel
                  + tooltb-rec(tbterm(action)) .
                    TKernel
                  + tooltb-rec(tbterm(process-status)) .
                    tooltb-snd(tbterm(process-status)) .
                    TKernel
                  + tooltb-rec(tbterm(start-process)) .
```

```
            TKernel
        + tooltb-rec(tbterm(quit))
  end TKernel
```

### 7.3 The ToolBus Application

We show how the processes for the tools are imported and put in parallel in the module SimulatorSystem.

```
process module SimulatorSystem
begin
   exports
   begin
      processes
         SimulatorSystem
   end
   imports
      .
      .
      .
      NewTool {
         Tool bound by [
            Tool → PT-Kernel
         ] to PKernel
         renamed by [
            TBProcess → Kernel
         ]
      },
      .
      .
      .
   definitions
      SimulatorSystem =
           Kernel
         ‖ StartProcess
         ‖ ActionChooser
         ‖ Function
         ‖ TraceCtrl
         ‖ BreakCtrl
         ‖ Display
end SimulatorSystem
```

And finally we put this in the ToolBus environment.

```
process module Simulator
begin
   imports
      NewToolBus {
         Application bound by [
            Application → SimulatorSystem
         ] to SimulatorSystem
         renamed by [
            ToolBus → Simulator
         ]
      }
end Simulator
```

### 7.4 Further Specification of the the Kernel Tool

We want to split the Kernel tool into a separate adapter and tool, so that a final implementation of the kernel can be used in other applications. We do this again by applying the refining and constraining techniques. We take the specification of the Kernel tool as given in section 7.2 and apply the following mapping, where the first rule is a default mapping.

```
tooltb-rec(tbterm($1))   →
                         tooltb-rec(tbterm($1)) .
                         tooladapter-snd($1)
action-choose-list    → tooladapter-rec(action-choose-list)
halt                  → tooladapter-rec(halt)
```

```
tooltb-rec(tbterm(process-status)) →
                    tooltb-rec(tbterm(process-status)) .
                    tooladapter-snd(process-status) .
                    tooladapter-rec(process-status)
```

By renaming TKernel into AKernel we obtain the adapter of the Kernel as shown below.

```
process module AKernel
begin
    exports
    begin
        processes
            AKernel
    end
    imports
        SimulatorData,
        ToolAdapterPrimitives,
        ToolToolBusPrimitives,
        ToolFunctions
    definitions
        AKernel =
              tooltb-rec(tbterm(compute-choose-list)) .
              tooladapter-snd(compute-choose-list) .
              (
                 tooladapter-rec(action-choose-list) .
                 tooltb-snd(tbterm(action-choose-list))
              + tooladapter-rec(halt) .
                 tooltb-snd(tbterm(halt))
              ) . AKernel
            + tooltb-rec(tbterm(action)) .
              tooladapter-snd(action) .
              AKernel
            + tooltb-rec(tbterm(process-status)) .
              tooladapter-snd(process-status) .
              tooladapter-rec(process-status) .
              tooltb-snd(tbterm(process-status)) .
              AKernel
            + tooltb-rec(tbterm(start-process)) .
              tooladapter-snd(start-process) .
              AKernel
            + tooltb-rec(tbterm(quit)) .
              tooladapter-snd(quit)
end AKernel
```

Now we specify the new Kernel tool.

```
process module TKernel
begin
    exports
    begin
        atoms
            snd : Tterm
            rec : Tterm
        processes
            TKernel
    end
    imports
        SimulatorData
    definitions
        TKernel =
              rec(compute-choose-list) .
              (
                 snd(action-choose-list)
              + snd(halt)
              ) . TKernel
            + rec(action) .
              TKernel
            + rec(process-status) .
              snd(process-status) .
              TKernel
            + rec(start-process) .
              TKernel
            + rec(quit)
```

    **end** `TKernel`

We constrain the adapter with the tool as follows.

```
process module TA-Kernel
begin
   imports
      NewToolAdapter {
         Tool bound by [
            tool-snd → snd,
            tool-rec → rec,
            Tool → TKernel
         ] to TKernel
         Adapter bound by [
            Adapter → AKernel
         ] to AKernel
         renamed by [
            ToolAdapter → TA-Kernel
         ]
      }
end TA-Kernel
```

And we change in the module PKernel the constraining by TKernel into TA-Kernel.

A generated animation of the complete specification of the simulator as ToolBus application is shown in figure 10.

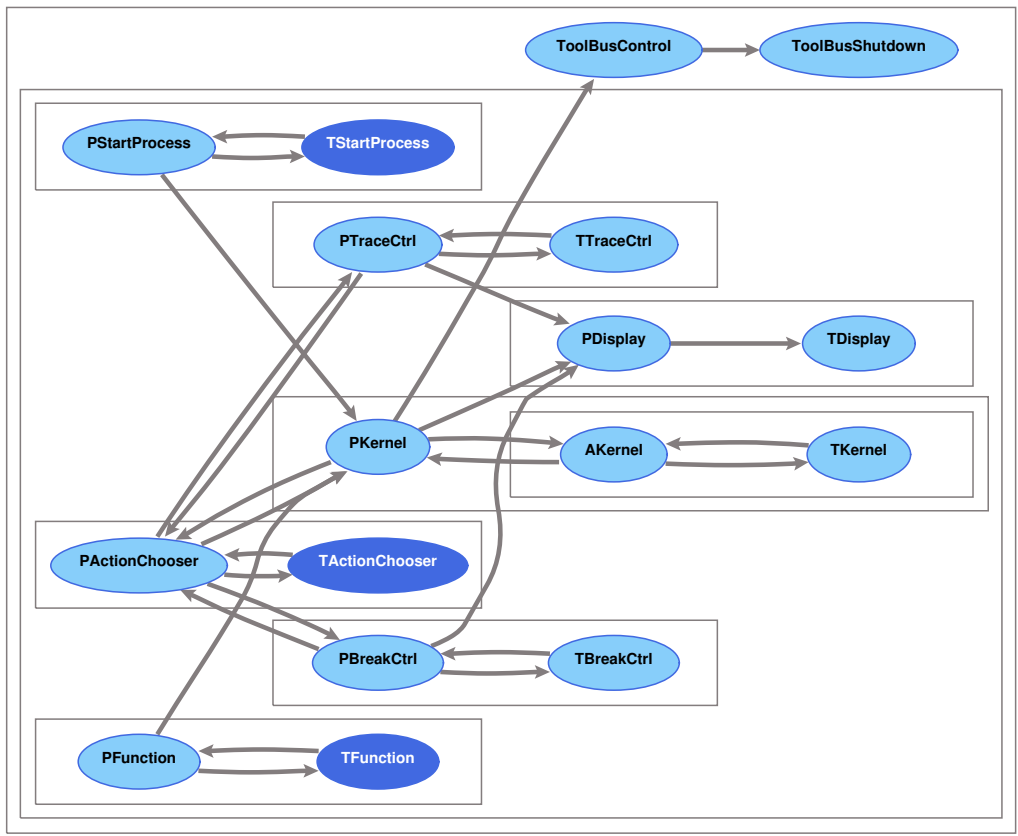

**Figure 10.** System design of the simulator[4]

---

4.    Generated with a left to right orientation instead of top to bottom

# 8. Implementation of the Simulator

The specification of the tools in the ToolBus application specification of the simulator is fine enough to proceed with the implementation of the simulator. Although the specification of the kernel is far too simple for such a complex tool, it is satisfactory here because we use the old simulator as base for the new implementation.

## 8.1 Kernel

Using the code of the old simulator as base we obtain an implementation of the kernel by doing the following

- remove the graphical user interface
- take out the embedded state machine
- add a component interface for communication with the outside world.

Of course the above three items are strongly related. An event comes from the gui and on handling may cause a change of state in the state machine.

In the implementation of the kernel an event comes through the component interface. This event is handled and if necessary a reply is send back through the component interface. The component interface really is an extension of the interface used in the coupling of the simulator with the animation. The function of the state machine is lifted from the kernel and is now served by the ToolBus.

The adapter of the kernel is implemented in Perl [35] on top of the general Perl adapter provided with the ToolBus. Perl is chosen because of its powerful regular expression matching and environment interaction.

## 8.2 Other Tools

The other tools are small and simple, and so they are easy to implement. We therefore do not give a further description of their implementation. We have chosen to implement them in Tcl/Tk, mainly because of the ease to build a gui within this language, and its widespread availability.

## 8.3 ToolBus Script

The ToolBus script for controlling the separate tools of the simulator can be derived from the ToolBus processes in the specification of the simulator as ToolBus application. This transformation is done by hand mainly because in the specification recursion is used to hold the state of a process and in a ToolBus script this has to be done with iteration and state variables.

## 8.4 Aggregation of Gui's

Except for the kernel, each tool has its own graphical user interface (gui), what looks rather shabby. So we like to integrate them into one big gui. In Tcl/Tk it is possible to indicate that a frame window is to serve as a container of another application and that a toplevel window is to be used as the child of such a container window. Following this scheme, we have implemented a separate tool that does the layout of several container windows. This layout can be resized as a whole and some windows can be resized in relation to each other through the use of paned windows.[5] A user preferring a different layout can implement another version similar to this.

For a toplevel to act as a child of a container window, it needs the window id of the parent. So the aggregated gui implementation has to communicate a window id to each child. The ToolBus script has been supplied with an initialization phase that receives all the id's of the container windows from the aggregated gui and distributes them over the tools. Each tool now first receives its parent id before doing

---

5. A paned window consists of two horizontal or vertical panes separated by a movable sash, and each pane containing a window.

anything else. The resulting gui is shown in figure 11.

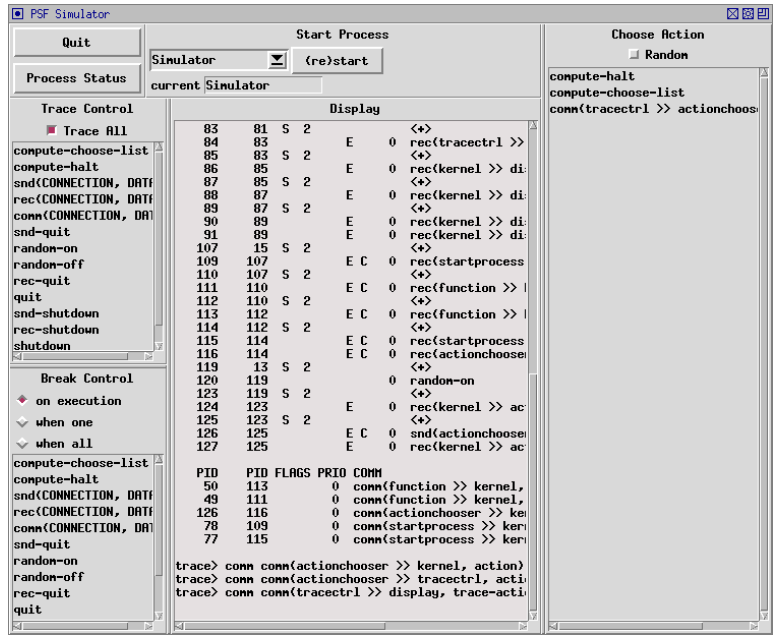

**Figure 11.** Aggregation of gui's

*8.5 Simulator*

To control the execution of the ToolBus we use a Perl script that sets up the environment in which the ToolBus and all the tools that make up our application run. This environment is needed to distribute arguments given on the command line to the different tools.

## 9. Extension with History Mechanism

In this section we describe the extension of the simulator with an history mechanism. The changes that have to be made to all levels of the design process are dealt with. This will show the impact of a software evolution process iteration on our design process.

*9.1 Architecture Specification*

The history actions consist of undo, redo, mark, and goto mark. The logical place for keeping an history is the kernel. We can let the kernel save the current state after every action it has done, but when running randomly this can use up a lot of memory and usually with an undo the user wants to jump directly to the state before random mode was started. Since the kernel does not know when the simulator is running randomly, it has to be informed when to save the current state. The action undo, redo, and goto mark, can all be seen as a goto to a certain state. So it suffices to add only a save and goto request to the kernel. Below we show the changes for the kernel with existing code in grey.

```
Kernel = Kernel(true)
Kernel(wait) =
  (
    [wait = false]→ (
      compute-choose-list .
      (
        action-choose-list .
        snd(kernel >> actionchooser, action-choose-list)
```

```
        +  halt .
           snd(kernel >> actionchooser, halt) .
           snd(kernel >> display, halt)
        )
    ) .
    Kernel(true)
 +  [wait = true]→ (
        .
        .
        .
        +  rec(actionchooser >> kernel, save) .
           Kernel(true)
        +  rec(actionchooser >> kernel, goto) .
           Kernel(false)
    )
)
```

Note that we also send a halt to the actionchooser now. Previously, in this case there was nothing to do for the actionchooser, but now a history action can take place.

A history action can be seen as just another action the user can choose from the all possible actions, so the logical place for such an action is in the actionchooser.

```
ActionChooser = Choose(false, false)
Choose(random, choose) =
      rec(kernel >> actionchooser, action-choose-list) .
      .
      .
      .
  +  rec(kernel >> actionchooser, halt) .
     force-random-off .
     Choose(false, true)
  +  [choose = true]→ (
        choose-action .
        snd(actionchooser >> kernel, action) .
        .
        .
        .
     +  snd(actionchooser >> kernel, save) .
        Choose(random, true)
     +  [random = false]→ (
           snd(actionchooser >> kernel, goto) .
           Choose(false, false)
        )
     )
  +  rec(kernel >> actionchooser, reset) .
     Choose(random, false)
  +  [random = true]→ (
        random-off .
        Choose(false, choose)
     )
  +  [random = false]→ (
        random-on .
        Choose(true, choose)
     )
```

We have to turn off the random mode on a halt so that a history action can be chosen. Note that the actionchooser can do a save also in random mode, what makes other history saving schemes possible, for instance every *n* steps.

### 9.2 ToolBus Application Specification

To obtain a ToolBus Application specification with added history mechanism from the architecture specification, we extend the mapping from section 7.1 with the following rules.

module Kernel

```
rec(actionchooser >> kernel, save)  →
                  tb-rec-msg(actionchooser, kernel, tbterm(save)) .
```

```
                        tb-snd-do(KERNEL, tbterm(save))
rec(actionchooser >> kernel, goto)  →
                        tb-rec-msg(actionchooser, kernel, tbterm(goto)) .
                        tb-snd-do(KERNEL, tbterm(goto))
```

module ActionChooser

```
    save              →  tb-rec-event(ACTIONCHOOSER, tbterm(save)) .
                         tb-snd-ack-event(ACTIONCHOOSER, tbterm(save))
    goto              →  tb-rec-event(ACTIONCHOOSER, tbterm(goto)) .
                         tb-snd-ack-event(ACTIONCHOOSER, tbterm(goto))
```

The adapter and the kernel tool can simply be extended with alternatives for handling a save and goto as follows.

```
        AKernel =
                  .
                  .
                  .
            +  tooltb-rec(tbterm(save)) .
               tooladapter-snd(save) .
               AKernel
            +  tooltb-rec(tbterm(goto)) .
               tooladapter-snd(goto) .
               AKernel

        TKernel =
                  .
                  .
                  .
            +  rec(save) .
               TKernel
            +  rec(goto) .
               TKernel
```

The adaptation of the actionchooser tool is slightly more complicated because of the distinguishing between the cases when there is a list of actions to choose from available and when there is not.

```
        TActionChooser = Choose(false)
        Choose(random) =
             tooltb-rec(tbterm(action-choose-list)) .
             (
                [random = true]→ (
                       .
                       .
                       .
                )
             + [random = false]→ (
                   tooltb-snd-event(tbterm(save)) .
                   tooltb-rec-ack-event(tbterm(save)) .
                   (
                      tooltb-snd-event(tbterm(random-on)) .
                      tooltb-rec-ack-event(tbterm(random-on)) .
                      tooltb-snd-event(tbterm(action)) .
                      tooltb-rec-ack-event(tbterm(action)) .
                      Choose(true)
                   +  tooltb-snd-event(tbterm(action)) .
                      tooltb-rec-ack-event(tbterm(action)) .
                      Choose(random)
                   +  tooltb-rec(tbterm(reset)) .
                      Choose(random)
                   +  History
                   )
                )
             +  [random = false]→ (
                    .
                    .
                    .
                )
```

```
                +  [random = true]→ (
                         .
                         .
                         .
                  )
                + tooltb-rec(tbterm(reset)) .
                  Choose(random)
                + [random = false]→ (
                    History
                  )
         History =
                tooltb-snd-event(tbterm(goto)) .
                tooltb-rec-ack-event(tbterm(goto)) .
                Choose(false)
```

The actionchooser tool only does a save when random mode is off, and so constrains the ToolBus process.

### 9.3 Implementation

In order to distinguish the different saves of history we need an unique id for every save. Then a goto send by the actionchooser can be supplied with an id so that the kernel can jump to the right saved history.

The actionchooser needs to generate these id's. We have implemented the id's as natural numbers and use ordering for easy lookups by the kernel. A mark of a saved history is done in the actionchooser by pairing this mark with the id of that save.

The history mechanism in the kernel is based on the history mechanism of the old simulator with only a few adaptations since some functionality is taken over by the actionchooser.

The gui of the history mechanism is implemented as a separate part of the actionchooser as shown in figure 12.

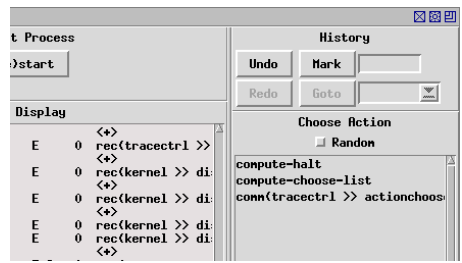

**Figure 12.** Aggregation of gui's with history

## 10. Coupling to Animation

The implementation of the old simulator coupled to the animation was done through the ToolBus as described in [13]. With that implementation the user could switch between choosing actions through the animation or from a list of actions. For our new implementation we have a choice from three possibilities:

- replacement of the actionchooser with the animation,
- use of two choosers controlled by the ToolBus,
- combination of the two choosers in one tool.

We choose to combine the two choosers, because both are implemented in Tcl/Tk and therefore the animation can be implemented as a toplevel window in the actionchooser with easy control of both choosers, and without change in the graphical user interface. And with this choice there is no need for adaptation of the architecture either.

## 11. Features not Implemented

Here we mention the features of the old simulator that are not implemented by the new simulator because they are seldom used. We give some indications on how these features can be implemented.

weighted random
: Normally all actions have equal chance to be picked randomly. With weighted random the position of an action in the process tree is taken into account. For instance, an action in parallel with a process that spans many actions sees its chance to be chosen reduced a lot by all these actions, but with weighted random its chance stays the same, and the actions of the parallel process get a combined weight equal to the weight of that one action.
This can easily be implemented by letting the kernel send weights with each action in the action-choose-list.

(re)load specification
: Because of the very short start up time of the old simulator, this feature is seldom used. The start up time of the new simulator does not differ much.
It can be implemented by letting the kernel do a clean up and start with a new specification or by shooting off the kernel and starting a new one.

trace to standard output / from standard input
: With trace to standard output every step of the simulator can be recorded and played back with trace from standard input. This can be used for demo's or for testing starting at a certain point every time, which also can be done with a mark on a saved history. Although these features are seldom used, they can be very convenient. Especially trace form standard input, because with that we can build applications with a stateless kernel for not too large simulations where a complete trace is fed to the kernel everytime together with a new action, such as a demo on the world wide web. So this probably will be implemented some time.
Trace to standard output can be implemented by embedding a monitor in the ToolBus that record all necessary actions, and trace from standard input can be put in place of the actionchooser.

## 12. Comparison of Implementations

In table 1 we compare the two implementation by lines of code. The new implementation takes considerably less lines of code mainly because Tcl/Tk and Perl code as TB scripts are very expressive, but it is also caused by the reduction of the complexity of the code. The left out features also play a role here but not by a large amount.

**Table 1.** Lines of code of the implementations

| language | lines of code | |
|---|---|---|
| | old | new |
| C | 21076 | 13884 |
| Tcl/Tk | | 1550 |
| Perl | | 179 |
| ToolBus script | | 281 |
| total | 21076 | 15894 |

The new implementation should be easier to maintain because of the reduction in lines of code and complexity, although it requires the knowledge of several more implementation languages. The specifications of the architecture and the simulator as ToolBus application play an important role here, since they can be used not only to get familiar with the design but also for testing changes and new features.

The graphical user interface has improved a lot, but it can also easily be altered. It should not be difficult to make an implementation that can be customized according to the preferences of each user.

The division in components has made reuse of parts of the implementation far more easier. It can even be used as a framework for simulation of other languages similar to PSF or new versions of PSF by only providing a different kernel.

The trade-off is that the new implementation is considerably slower, about a factor of thirty. This is due to the fact that this implementation consists of many processes running at the same time and the inter-process communications take up a lot of time. For working interactively this is not a problem, but for large random simulations, for instance validation testing, it is too slow.

**13. Related Work**

In literature several architecture description languages have been proposed and some are based on a process algebra, such as *Wright* [2], *Darwin* [20], and *PADL* [7]. A comparison of several ADL's can be found in [23]. Most of the ADL's do not have any or very little support for refinement. SADL [25][26] however, has been specially designed for supporting architecture refinement. In SADL, different levels of specifications are related by refinement mappings, but the only available tool is a checker.

Formal development techniques such as B [1], VDM [16], and Z [10] provide refinement mechanisms, but they do not have support for architecture descriptions. The $\pi$-Method [27] has been built from scratch to support architecture-centric formal software engineering. It is based on the higher-order typed $\pi$-calculus and mainly built around the architecture description language $\pi$-ADL [28] and the architecure refinement language $\pi$-ARL [29]. Tool support comes in the form of a visual modeller, animator, refiner, and code synthesiser.

LOTOS [8], a simular specification language to PSF, is used in [17] for the formal description of architectural styles as LOTOS patterns, and in [33] it is used as an ADL for the specification of middleware behaviour.

**14. Conclusions**

The development of the architecture of the simulator in the form of a specification in PSF turned out very well. We were able to start with a simple architecture and extend it with more functionality without any difficulties. The transition from architecture to system design in the form of a ToolBus application specification by means of vertical and horizontal implementation proved to be succesful. The extension with the history mechanism showed that adding functionality to a finished product did not lead to any problems in the software development process. The PSF Toolkit played an important role. The simulation and animation provided a good view of the behavior of the specifications. A change in a specification could be tested on the fly because of the automatic generation of animations. These animations can be very useful for someone who has to adapt the software product and who is not familiar with it. The animation of the architecture can also be used for communicating the design to the stakeholders in the development process.

The implementation of the simulator has improved a lot, especially its interface. The coupling with animation is smoother since in the old situation there was one chooser from the simulator and one from the animation and a switch over was needed to use the other chooser, now the two choosers are integrated in one tool and can be used simultaneously. The maintainability of the simulator has increased caused by the division into components and the reduction in complexity, but mostly by the specification of the architecture and system design. Although the new implementation is much slower, it still has a good performance when working interactively and for small random simulations.

Future work may concentrate on other system design models than the ToolBus. Here we did not do any refining of the specification of the components, since they were not useful here. But such refinements may make use of certain styles or patterns and be applied on different levels of the design. Although the tools from the PSF Toolkit were sufficient for the work we have done sofar, future work may ask for more support. We think of a tool for the automic applications of mappings. Here we used an ad hoc tool only for checking of the mappings.